# Linking studies of tiny meteoroids, zodiacal dust, cometary dust and circumstellar disks

A.C. Levasseur-Regourd[1], C. Baruteau[2], J. Lasue[2], J. Milli[3,4], J.-B. Renard[5]

**Addresses**
[1] LATMOS, Sorbonne Université, CNRS, CNES, Paris, France
[2] IRAP, Université de Toulouse, CNES, CNRS, UPS, Toulouse, France
[3] European Southern Observatory, Alonso de Córdova 3107, Casilla 19001, Santiago, Chile
[4] Univ. Grenoble Alpes, IPAG, Grenoble, France
[5] LPC2E, Université d'Orléans, CNRS, Orléans, France

**Key words:** meteoroids; dust particles; zodiacal cloud; comets; circumstellar disks; light scattering

**Abstract**

Tiny meteoroids entering the Earth's atmosphere and inducing meteor showers have long been thought to originate partly from cometary dust. Together with other dust particles, they form a huge cloud around the Sun, the zodiacal cloud. From our previous studies of the zodiacal light, as well as other independent methods (dynamical studies, infrared observations, data related to Earth's environment), it is now established that a significant fraction of dust particles entering the Earth's atmosphere comes from Jupiter-family comets (JFCs).

This paper relies on our understanding of key properties of the zodiacal cloud and of comet 67P/Churyumov-Gerasimenko, extensively studied by the Rosetta mission to a JFC. The interpretation, through numerical and experimental simulations of zodiacal light local polarimetric phase curves, has recently allowed us to establish that interplanetary dust is rich in absorbing organics and consists of fluffy particles. The ground-truth provided by Rosetta presently establishes that the cometary dust particles are rich in organic compounds and consist of quite fluffy and irregular aggregates. Our aims are as follows: (1) to make links, back in time, between peculiar micrometeorites, tiny meteoroids, interplanetary dust particles, cometary dust particles, and the early evolution of the Solar System, and (2) to show how detailed studies of such meteoroids and of cometary dust particles can improve the interpretation of observations of dust in protoplanetary and debris disks. Future modeling of dust in such disks should favor irregular porous particles instead of more conventional compact spherical particles.

## 1. Introduction

*1.1. Concise historical background*

While human beings of all times have certainly noticed meteor showers, significant progress about the sources of these phenomena has taken place since the 19th century (for reviews, Jenniskens 2006; Koschny et al. 2019). It was understood that they originate from cosmic particles of sizes below hundred of micrometers, here named tiny meteoroids, entering the Earth's atmosphere on parallel trajectories. Giovanni Schiaparelli and Urbain Le Verrier found similarities between their orbits and those of periodic comets, now named Jupiter-family comets (JFC). Typical examples are the Perseids and comet 109/Swift-Tuttle, and the Leonids and comet 55P/Tempel-Tuttle. Comets are indeed classified from their dynamical properties. JFCs move on direct orbits with periods below 20 years, rather low inclinations on the ecliptic and moderate aphelion distances, because of previous gravitational perturbations by Jupiter, while HTCs, for Halley-Type Comets, have periods between 20 and 200 years and may present any inclination. Also, comets on nearly parabolic orbits and having periods above 1000 years probably come from the Oort cloud; they are named OTCs for Oort-Type Comets or, alike, OCC for Oort-Cloud Comets. While initial observations suggested comets to be somehow "sand banks" held together by gravity, the model proposed by Fred Whipple for non-gravitational forces predicted the existence of a cohesive nucleus inside the gas and dust coma (Whipple 1950). The space exploration of comets, beginning with the flybys of 1P/Halley in 1986, soon established the existence of low-density nuclei (for a review, Festou et al. 2004). Such missions have begun to provide a wealth of information on cometary dust, and indirectly on tiny meteoroids.

*1.2. Interplanetary dust particles, origins and evolution*

Small dust particles (with sizes about a few µm to hundreds of µm), once ejected from the nucleus of an active comet, may form spectacular cometary dust tails, which point away from the nucleus in the anti-solar direction under the effect of the solar radiation pressure. Larger dust particles (with sizes about 1 mm to at least a few dm) are also detected in the infrared domain (e.g., Reach & Kelly 2007), near the orbits of comets. These so-called cometary dust trails arise from dust particles that remain close to their parent bodies because they are not significantly affected by perturbations (Agarwal et al. 2010; Soja et al. 2015). Upon entering the Earth's atmosphere, tiny meteoroids of cometary origin induce periodic meteor showers, which become permanent after the dust particles have been scattered all over the stream under perturbations.

Within the Solar System, some dust particles may come from comets, asteroids, the environment of giant planets, and the interstellar medium (for reviews, Grün et al. 2001; Hajduková Jr. et al. 2019). All together, they contribute to the formation of a huge and flattened circumstellar cloud, the interplanetary dust cloud, also named zodiacal cloud because it is detectable through the zodiacal light, i.e. the solar light scattered by interplanetary dust particles (for a review, Levasseur-Regourd et al. 2001). The major sources of interplanetary dust are the above-mentioned active cometary nuclei, releasing dust and gases through sublimation of their ices, and the asteroids, likely to suffer collisions within the main asteroid belt. The discoveries by infrared space telescopes of (i) near-ecliptic zodiacal dust bands associated with collisions between, e.g., asteroids of Themis, Koronis, Beagle, Karin or Veritas families (e.g., Dermott et al. 1984; Nesvorný et al. 2008), (ii) of cometary dust trails (e.g., Agarwal et al. 2010), have been of major importance to emphasize the contribution of asteroids and comets to the zodiacal light.

The interplanetary dust particles are not only affected by the solar gravity field. For instance, very small and fluffy dust particles are blown away by the solar radiation pressure on hyperbolic orbits, while small dust particles slowly spiral towards the Sun under the Poynting-Robertson drag (for a review, Vaubaillon et al. 2019). Sources and sinks thus exist within the interplanetary dust cloud, which is far from being homogeneous. This was suggested in the past by zodiacal light observations (Levasseur & Blamont 1973). It is nowadays perfectly illustrated by images of the sky thermal emission, providing evidence for cometary trails, and asteroidal debris with toroidal distributions in the asteroidal belt (Levasseur-Regourd & Lasue 2019, Fig. 4).

Section 2 presents clues, obtained from independent approaches, on the cometary origin of most near-Earth tiny meteoroids. Section 3 summarizes the properties of cometary dust particles, as revealed by the Rosetta cometary mission, before mentioning their significance, back in time and beyond in space. Section 4 analyzes the implications of the previous results for remote observations of circumstellar disks, for which the light scattering properties of dust particles could be better explained by more realistic models that include porous irregular aggregates, rather than compact spheres.

**2. Comets as the main source of tiny meteoroids entering the Earth atmosphere**

The dust concentration within the interplanetary dust cloud grows both towards its near-ecliptic symmetry plane and towards the Sun. Interestingly enough, different approaches have progressively established that most of the near-Earth interplanetary dust particles that may enter the Earth's atmosphere are of cometary origin.

*2.1. Clues from zodiacal light studies*

Remote observations of the zodiacal light provide information on the physical properties of the interplanetary dust particles (for a review, Lasue et al., this issue). Once observations of the linear polarization degree of the zodiacal light along a given line of-sight are tentatively inverted, at least in the near-ecliptic symmetry plane of the zodiacal cloud, phase curves may be retrieved at 1.5 au solar distance. The dependence of the polarization degree at 90° phase angle is also monitored from 1.5 au down to the solar F-corona (Levasseur-Regourd et al. 2001). The variations with phase angle and solar distance can be interpreted through numerical models and experimental simulations, in the laboratory and under reduced gravity conditions (for reviews, Lasue et al. 2015; Levasseur-Regourd et al. 2015).

The overall shape of the local polarimetric phase curves, which are smooth and feature a negative branch in the backscattering region (Fig.1), suggests that the scattering dust particles are not spherical. At 1.5 au from the Sun near the ecliptic, we have obtained a good fit of the local polarization degree by assuming irregular spheroids and fractal aggregates thereof, and complex refractive indices representative of silicates and more absorbing refractory organics; they would correspond to a contribution of dust particles of cometary origin above 20% in mass (Lasue et al. 2007). Satisfactory experimental data points were also obtained for mixtures of silicates and organics forming compact and fluffy particles, with (40 ± 5)% of absorbing organics at 1.5 au (Hadamcik et al. 2018).

Furthermore, out of ecliptic observations have been tentatively inverted (Renard et al. 1995). They suggest the existence of at least two populations of dust, with different average orbital inclinations, originating from periodic comets and from new comets.

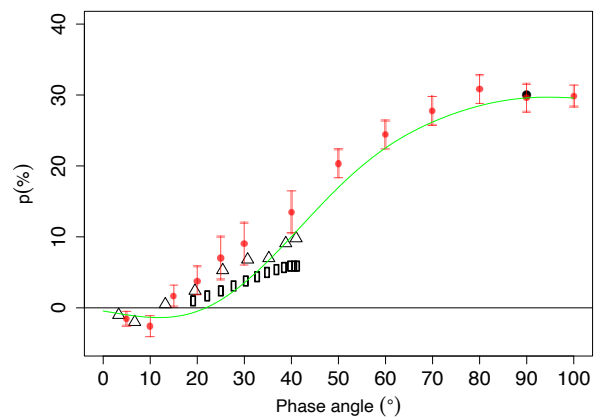

**Figure 1.** Local polarization degree as a function of the phase angle at 1.5 au from the Sun near the ecliptic. Open circles and triangles correspond to an inversion by the nodes of lesser uncertainty method (Dumont & Levasseur-Regourd 1985) of, respectively, observations compiled from various authors (Fechtig at al. 1981) and observations (Dumont & Sanchez 1975) avoiding contamination from atmospheric airglow. The black dot at 90° results from a rigorous inversion on the Earth's orbit, tentatively extrapolated at 1.5 au (e.g., Levasseur-Regourd et al. 1999). The green curve corresponds to the best numerical fit (Lasue et al. 2007). The red data points correspond to the best experimental mixture of minerals and carbonaceous compounds (Hadamcik et al. 2018).

The variation with solar distance R of the local linear polarization degree at 90° in the symmetry plane follows a power law about [(30 ± 3) $R^{+0.5 \pm 0.1}$]% between 0.3 and 1.5 au (Fig. 2). The decrease in polarization with decreasing solar distance suggests a change in dust properties with increasing temperature (Levasseur-Regourd et al. 2001). Fitting theoretically the solar distance dependence provides clues to a progressive decrease of organics with decreasing solar distance (Lasue et al. 2015). Experimental simulations on clouds of dust particles are consistent with a constant ratio of (35 ± 10) % in mass of fluffy aggregates versus compact particles, and a decreasing ratio of organics with decreasing solar distance (Hadamcik et al. 2018). The ratio of organics, estimated to be close to 40% at 1 au, is less constrained at 1.5 au, where it could be in a 60 to 50% range.

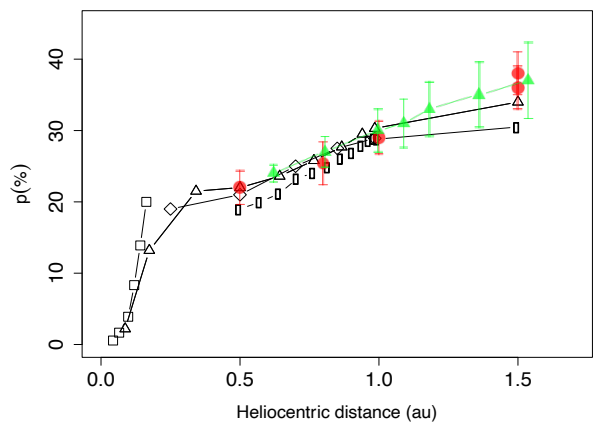

**Figure 2.** Local polarization degree at 90° phase angle as a function of solar distance near the ecliptic (error bars below 10%) Open circles and triangles correspond to an inversion of observations compiled from various authors (Fechtig at al. 1981) and of observations that avoided contamination from atmospheric airglow (Dumont & Sanchez 1975). Open squares confirm from a different approach (Mann et al. 1990) the existence of a drastic change at low solar distances.
The green line corresponds to a numerical model with decreasing values of organics, e.g. 40% at 1 au and 0% at 0.5 au (Lasue et al. 2007). The red dots correspond to the best experimental mixtures, with decreasing percentages in carbonaceous compounds, about 60, 40, 30 and 0%, respectively at 1.5, 1, 0.8 and 0.5 au (Hadamcik et al. 2018).

Such properties agree fairly well with what has been estimated for those of cometary dust particles (see Subsection 3.1), which are now understood to be rich in minerals and organics. Indeed, complex refractory organics are progressively sublimating from solid to gas phase when getting closer to the Sun.

*2.2. Clues from other studies*

An elaborate zodiacal cloud model, based on the orbital properties and lifetimes of comets and asteroids, and on the dynamical evolution of dust after its release, has been developed by Nesvorný et al. (2010). The model, which is constrained by IRAS observations, is consistent with other observations of the zodiacal cloud and meteors, as well as with spacecraft impact experiments and properties of collected micrometeorites. The authors conclude that the dust particles produced by JFCs represent about 85% in mass of the total influx on the Earth's atmosphere. This flux of micrometeorites has been estimated to be about (30 ± 20) x $10^6$ kg/yr (Plane et al. 2012).

A model based on IRAS and COBE infrared observations has been proposed by Rowan-Robinson & May (2013). It leads to the conclusion that, at 1 au from the Sun in the near-ecliptic symmetry surface, comets could contribute to 60-80% of the zodiacal cloud, with asteroidal and interstellar dust particles contributing together to the remaining fraction.

Finally, a more indirect approach, relying mainly on the modeling of metal atoms layers in the Earth's mesosphere and on data related to cosmic spherules accretion rate in Antarctica, has estimated that JFCs contribute to (80 ± 17) % of the total mass of dust in the terrestrial atmosphere (Carrillo-Sánchez et al. 2016). As discussed by the authors, this result agrees with recent observations of the zodiacal thermal emission.

This converging trend, which is obtained by four different data sets and approaches, leads us to conclude that cosmic dust particles located at 1 au near the ecliptic mostly come from JFCs, and that they can be better defined by considering the properties of dust particles released by such comets.

## 3. Properties of near-Earth tiny meteoroids, a post Rosetta approach

The dust particles ejected from the nucleus of comet 67P/Churyumov-Gerasimenko (thereafter 67P/C-G), indeed one JFC, have been studied from the ESA Rosetta rendezvous spacecraft, which orbited the nucleus of 67P/C-G in 2014-2016, from 13 months before its perihelion passage to 13 months after it, as reviewed in Levasseur-Regourd et al. (2018).

*3.1. Composition and physical properties of cometary dust particles*

Composition and physical properties have been mostly monitored on Rosetta (i) by COSIMA, which has obtained microscopic images of collected dust particles before analyzing their composition with a mass spectrometer (e.g., Langevin et al. 2016; Fray et al. 2017), (ii) by MIDAS, which has imaged in 3D the surfaces of collected dust particles with its atomic force microscope (e.g., Bentley et al. 2016; Mannel et al. 2019), and (iii) by GIADA, which has measured the optical cross-section, speed and momentum for particle sizes of about 0.1 to 1 mm, and the cumulative flux of dust particles smaller than 5 µm (e.g., Della Corte et al. 2015; Fulle et al. 2016). The on-board OSIRIS cameras, which monitored the trajectory of large ejected particles (e.g., Ott et al. 2017), and COSAC, which analyzed outgassing particles by mass spectrometry on board the Philae lander (e.g., Goesmann et al. 2015), have also provided unique results about the properties of dust particles within 67P/C-G coma.

The elemental and isotopic compositions of dust particles have been analyzed; their mineralogy and organics composition have been extensively studied. We stress that the refractory organics phase is dominated by high-molecular weight components, and that comets are an important reservoir of carbon and organic matter, as established from various approaches (e.g., Herique et al. 2016; Fray et al. 2017).

Cometary dust particles are aggregates or agglomerates of grains (down to sizes about 100 nm and even less), with hierarchical structures and fractal dimensions going down to 1.7 (Güttler et al. 2019; Mannel et al. 2018). Morphologies of the particles range from very porous to quite compact, with volume filling factors covering many orders of magnitude. Densities are quite low, from a few tens to hundreds of kg/m$^3$. This is in good agreement with the estimation, about 100 kg/m$^3$ made for dust in the coma of comet Halley (Fulle et al. 2000), and even lower than the bulk density, of 537.8 ± 0.6 kg/m$^3$, of the nucleus of 67P/C-G (Pätzold et al. 2019).

Finally, in the inner coma, the dust-brightness phase curves reveal a flattened u-shape that may rule out a sharp surge in the forward-scattering region (Bertini et al. 2017; Fulle et al. 2018). Numerical as well as experimental simulations suggest that such trends also imply a significant amount of organic compounds and of fluffy aggregates (Moreno et al. 2018; Markkanen et al. 2018; Levasseur-Regourd et al. 2019).

It may be added that the properties of dust particles in the coma of 67P/C-G fairly agree with the results tentatively derived from remote light scattering observations of cometary and interplanetary dust (e.g., Levasseur-Regourd et al. 2007; Hadamcik & Levasseur-Regourd 2009).

*3.2. Comparison with cosmic dust particles collected in the stratosphere or in Antarctica*

A result of major importance for studies of tiny meteoroids is certainly the evidence for remarkable similarities found between cometary dust particles, with emphasis on 67P/C-G, and some interplanetary dust particles collected in the Earth stratosphere (IDPs) and micrometeorites (MMs) collected in central Antarctica snows.

Within the stratosphere, CP-IDPs (i.e., chondritic-porous IDPs) are optically-black and organics-rich particles consisting of porous dust aggregates. Similarities with dust samples, collected at high velocity during Stardust flyby of JFC comet 81P/Wild 2 (e.g., Hörz et al. 2006; Zolensky et al. 2008), have been confirmed by the Rosetta ground-truth, in terms of typology, elemental composition and mineralogy. Meanwhile, chondritic-smooth IDPs, representing about half of the collected IDPs, have been processed by aqueous alteration. They have been suggested to come from asteroids, or from main belt comets and active asteroids. UCAMMs (i.e. ultra carbonaceous Antarctica MMs) also present similarities with cometary dust particles, as established for 67P/C-G. They typically comprise more than 80% of carbonaceous material in volume, with substructures going down to about 50 nm in size (e.g., Nakamura et al. 2005; Engrand et al. 2016).

All together, it is still difficult to estimate the fraction of IDPs coming from comets and asteroids. An automatic classification of a few hundreds of cosmic dust particles, relying on their X-ray spectra, has nevertheless suggested comparable percentages (Lasue et al. 2010). Estimating a percentage may be even more difficult for Antarctica MMs, taking into account the small amount and the fragility of un-melted collected particles. The similarities between CP-IDPs or UCAAMs and cometary dust particles may be relatively easy to explain, once the orbits and the properties of the dust particles likely to enter the Earth atmosphere are considered. First, dust particles ejected from JFCs move close to their parent body on direct orbits, with rather low inclinations on the ecliptic and periods of less than 20 years, leading to moderate aphelion distances. Their relative velocities at atmospheric entry are thus below 15 km/s for particles in the 100s µm size range, that is to say significantly below those of HTCs and OTCs (Nesvorný et al. 2010). Secondly, their morphologies and porous structures allow their temperatures to remain low enough so that significant amounts of organics may survive in the atmosphere; fluffy aggregates may indeed bring up to $\pi^3$ times more material in volume without being ablated to the Earth's surface than compact spherical particles (see Fig. 10, Levasseur-Regourd et al. 2018).

*3.3. Significance of such results, back in time*

The aforementioned results provide clues to the formation of the Solar system. They suggest that the cometary dust particles were built in the external regions of the protosolar disk, from submicron-sized grains accreted at low collision velocities, with possible further addition of minerals processed close to the proto-Sun (e.g., Engrand et al., 2016; Blum et al. 2017; Fulle and Blum 2017; Lasue et al. 2019).

Such results are also of importance for a better understanding of the evolution of the Solar System at the late heavy bombardment epoch. By then, outer planets scattered a large number of cometary nuclei in the inner Solar System. These nuclei became active and ejected a huge amount of dust particles, possibly leading to a zodiacal light about $10^4$ brighter than nowadays (Booth et al. 2009; Nesvorný et al. 2010). An enormous incoming flux of interplanetary dust particles on the atmospheres of telluric planets might then have provided a massive delivery of pristine carbonaceous compounds on the young Earth, not so much before the emergence of early life. Those are merely speculations. It is nevertheless true that the study of tiny meteoroids brings us back in time, to the zodiacal dust cloud, to the comets, and to the Solar System early formation, and beyond in space, as discussed in the next section.

## 4. Implications for studies of protoplanetary and debris disks

*4.1 From the Solar System to further away in space*

The structure of exozodiacal dust clouds, driven by the orbits and masses of their exoplanets, is likely to be different from the structure of our zodiacal cloud. However, exocomets have been detected around many stars and thought to be pristine objects (e.g. Rappaport et al. 2018), not to mention objects that might be exocomets crossing the Solar System, such as 2I/Borisov (Guzik et al. 2019). It might be assumed that such comets were formed under conditions similar to our comets, and that dust particles present in circumstellar disks could be irregular porous aggregates, suggesting that their optical properties should not be modeled with compact spherical particles.

Protoplanetary disks are gas- and dust-rich disks surrounding young stars, and the seat of early planet formation and evolution. They evolve towards debris disks around main-sequence stars, after their gas content has been dispersed and partly incorporated into planetary atmospheres (for a review, Williams & Cieza 2011). Debris disks encompass planetesimals, as well as smaller dust particles.

*4.2. Protoplanetary disks*

The classical picture of protoplanetary disks being smooth, continuous structures has been challenged by the growing number of spatially resolved observations (e.g., Long et al. 2018). Radial discontinuities and large-scale asymmetries are common features of the emission of protoplanetary disks. It is the case for instance of the disk around the pre-main-sequence star MWC 758, which displays multiple spirals and arcs in near-infrared scattered light (e.g., Benisty et al. 2015), and asymmetric bright rings outside an approximately 40 au-wide cavity in the continuum emission at (sub-) millimeter wavelengths (Dong et al. 2018; Casassus et al. 2019), as illustrated in Fig. 3, left panel.

Disk sub-structures have stimulated a body of modeling works, in particular to determine whether they could be signatures of the presence of (unseen) planetary companions. Such modeling works usually employ radiative transfer calculations based on the results of gas and dust hydrodynamical simulations. The physical properties of the dust, like its size distribution, porosity and composition, are nevertheless poorly constrained in protoplanetary disks. Models of protoplanetary disks always nearly assume that dust has a power-law size distribution, is comprised of spherical compact particles with an internal mass volume density of a few 1000s kg/m$^3$, and has a mixed composition (silicates, amorphous carbons, water ices etc.).

Departure from this set of generic assumptions is rare, but we note that the disk model of Baruteau et al. (2019) can better reproduce the (sub)-millimeter observations of the MWC 758 disk if assuming moderately porous dust, with an internal density of around 100 kg/m$^3$ for dust particles between a few tens of microns and a centimeter in size (Fig. 3, middle panel). This rather low density is overall consistent with the ground-truth provided by Rosetta of porous aggregates in the inner coma of comet 67P/C-G. This comparison stresses that studies of the physical properties of cometary dust can greatly help constrain dust models of protoplanetary disks.

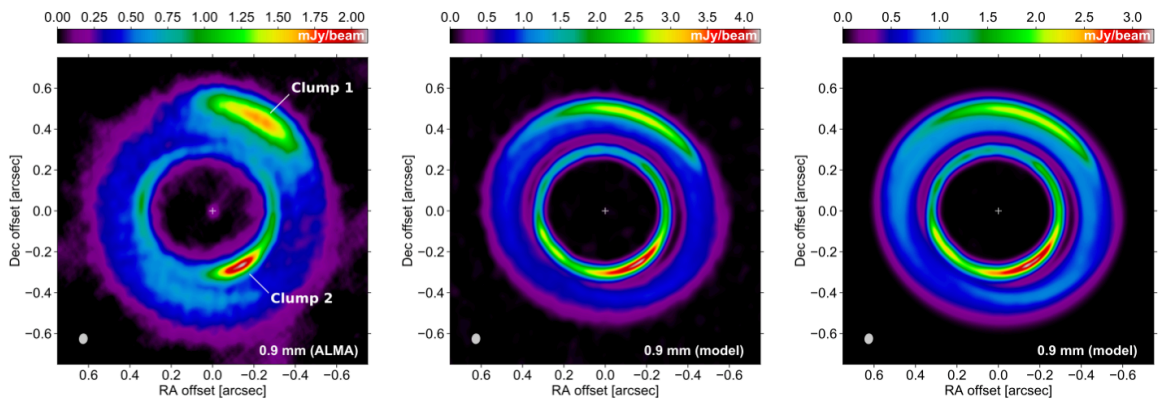

**Figure 3.** Observed and synthetic continuum emission at 0.9 mm wavelength in the protoplanetary disk around MWC 758. **Left**: ALMA observation (adapted from Dong et al. 2018). **Middle**: synthetic flux map (Baruteau et al. 2019) from an hydrodynamical model with two giant planets that structure the disk, and which features spherical porous dust particles (see text). **Right**: same synthetic map, but for a new dust radiative transfer calculation that uses a scattering phase function which fits that of the dust coma of comet 67P/C-G, measured by Rosetta/OSIRIS, and is, overall, in better agreement with the ALMA observations.

There is much room for improving dust models of protoplanetary disks, and we only list here a few salient points. One is to go beyond spherical dust particles, which would be particularly relevant for fluffy dust aggregates below hundreds of microns in size, for which the effective cross section of interaction with gas or radiation can be quite different from the surface area of a sphere. It demands to compute the dust's optical properties without resorting to Mie theory. It also demands new prescriptions for the aerodynamical drag between gas and (non-spherical) dust particles in hydrodynamical simulations of protoplanetary disks. Another area where progress can be made is in the treatment of dust scattering in synthetic observations. At (sub-) millimeter wavelengths, dust scattering is often ignored while it can largely reduce the emission in the optically thick parts of protoplanetary disks (Zhu et al. 2019). When dust scattering is accounted for, it is modeled either by an isotropic phase function or by an anisotropic phase function that only includes forward scattering. Neither phase function is actually consistent with that of the dust coma of comet 67P/C-G, which features both moderate backward and forward scattering at optical and near-infrared wavelengths (Bertini et al. 2017; Bockelée-Morvan et al. 2019). Adopting this observed phase function to the MWC 758 disk model of Baruteau et al. (2019) leads to a lower level of flux and an enhancement of the disk emission between the two asymmetric bright rings (Fig. 3, right panel). Although this result is still preliminary, it is interesting to note that the use of the phase function observed by Rosetta leads to a better agreement between the predicted and observed maps of continuum emission of the MWC 758 disk.

*4.3. Debris disks*

While using realistic scattering properties of dust particles measured in Solar System comets to compute the radiative transfer in protoplanetary disks is a first step towards more accurate models of the primordial stages of planetary systems, we need in parallel to understand the variety of dust particles properties that exists in those systems, to put our own Solar System in context. This can be done by measuring remotely and comparing the dust properties in debris disks. These disks represent ideal targets in this prospect, because they are optically thin, unlike protoplanetary disks. This means that the scattering properties of the dust can be directly retrieved from the analysis of resolved images of those disks at near-infrared/optical wavelengths.

Debris disks are found around at least 20% of nearby main-sequence stars in far-infrared surveys, and explained by the steady-state collisional erosion of planetesimal belts. They can be considered as a component of planetary systems, such as the Kuiper belt in our Solar System, and as such provide valuable information on the outcome of planet formation (for a review, Hughes et al. 2018). Unlike protoplanetary disks, the dust detected in those systems is of secondary origin and can be explained by collisions between planetesimals.

To date, only very massive analogs to our Kuiper belt, around young stars between 10 to 100 Myrs, have been angularly resolved in scattered light, mainly due to the technical difficulty to reach the required contrast and angular separation. In those systems, one or several rings of planetesimals produce smaller bodies through collisions, until the smallest dust particles (a few µm for A-type stars) leave the system, blown away by the stellar radiation pressure. Scattered light imaging in the near-infrared or optical is sensitive to these smallest dust particles.

The analysis of the photometry of a ring, in un-polarized or linearly polarized light, provides insightful properties of the dust, such as the brightness and linear polarization as a function of the phase angle (e.g., Hughes et al. 2018). These measurements then can be used to constrain the size, shape and composition of the dust particles.

Similarly to what was noticed for the interplanetary and cometary dust, and more precisely for comet 67P/C-G, the phase function is far from that predicted by the Mie theory for compact spheres. Interestingly, two systems (HD 35841 and HD 19089) display a phase function surprisingly similar to the flattened u-shape from the 67P/C-G extracted from the OSIRIS camera onboard Rosetta (Fig. 4). In both cases, fits assuming compact spheres following the Mie theory or even its variation, the Distribution of hollow spheres (Min et al. 2005), are not satisfactory. Comparing the properties of the dust in these two systems with those from the particles directly detected by Rosetta within the dust coma of the comet (see Subsection 3.1) can therefore yield insights into the particle shape, composition and size. This empirical comparative approach to the interpretation of the scattering properties of debris disks is highly complementary to the modeling approach currently applied, which reaches its limit especially when the scatterers are large irregular particles with respect to the wavelength still difficult to model numerically (Kolokolova et al. 2004).

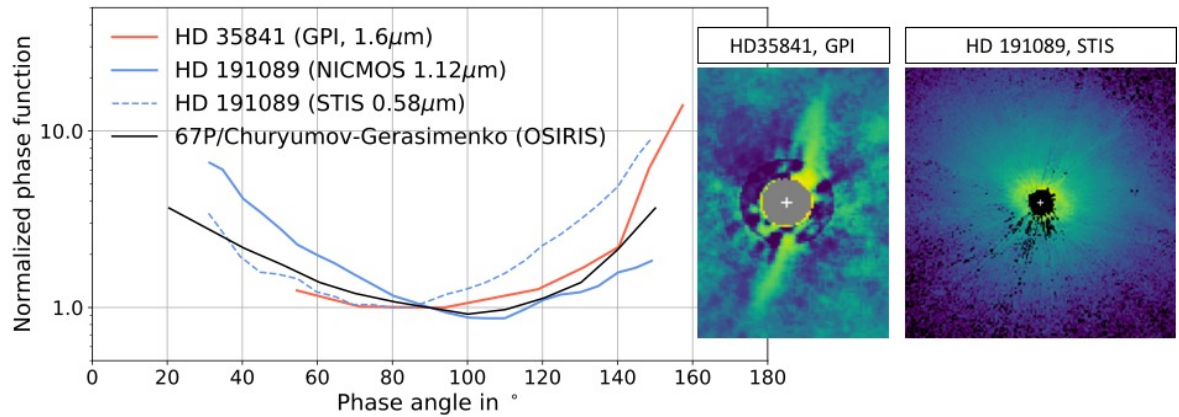

**Figure 4**. Comparison between brightness phase dependence for two debris disks and cometary dust. **Left.** Phase curves of two debris disks around HD35841 (measured in the H-band with the Gemini Planet Imager, Esposito et al. 2018) and HD191089 (measured in the optical with HST/STIS and in the J band with HST/NICMOS, Ren et al. 2019), presenting quite fair an agreement with those of 67P/C-G extracted from OSIRIS on 28 August 2015, soon after perihelion passage (Bertini et al. 2017). **Right.** Images of both debris disks.

The analysis of the polarization degree from debris disks and its comparison to results about cometary and interplanetary dust can also shed light on the properties of the dust in exoplanetary systems. For most systems, only the extraction of polarized light, that is to say the product of the normalized brightness by the linear polarization degree, is feasible. Whenever the linear polarization degree, which is difficult to retrieve, can be extracted, it is obtained over a wide range of phase angles, in contrast to comets where observations are seldom available beyond 90º.

Thanks to the favorable inclination, morphology and brightness of the HR4796 system, the polarized light phase angle dependence could be extracted from 14º to 166º in the optical. Its shape could be compatible with a negative branch of polarization below 20º, as shown in Fig. 5.

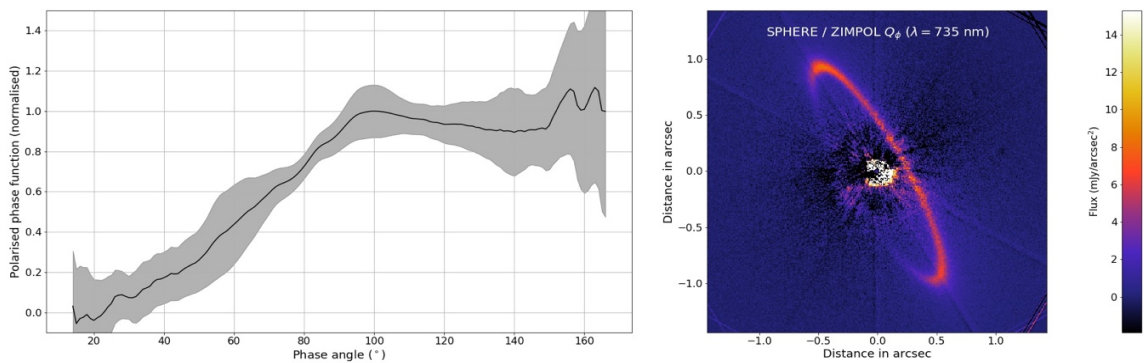

**Figure 5**. **Left**: Polarized light phase angle dependence from the debris dust around HR4796, averaged between the east and west sides of the disk. **Right**: Image of the debris disk around HR4796, obtained with VLT/SPHERE in the optical (600-900 nm) in linearly polarized light (North is up, East to the right). Adapted from Milli et al. (2019).

The maximum degree of polarization observed in debris disks could be relatively high, in the range 20 to 50% (e.g., Graham et al. 2007). As far as comets are concerned, although observations at large phase angles are often impossible to obtain, it is of interest to notice that i) the maximum in polarization is also by 100° phase angle and ii) for Oort-cloud comet 1995 O1/Hale-Bopp, which has presented the highest polarization degree ever measured (Hadamcik & Levasseur-Regourd 2003), an extrapolation of phase dependence could lead to a maximum in the 30 to 40% range. Such levels of polarization are not uncommon for aggregated particles with individual monomers smaller than the wavelength, as proposed in those debris disks systems (Milli et al. 2017).

*4.4. Discussion*

Linking different methodologies and fields is a difficult exercise, and some strengths and weaknesses of this approach are discussed below. We have provided robust evidence that comets are, within the Solar System, the main sources of tiny meteoroids entering the Earth's atmosphere; these organics-rich dust particles are found to be irregular and to present morphologies ranging from very porous to quite compact. However, extensive information about the local properties of interplanetary dust outside the symmetry plane of the zodiacal cloud is still missing. Also, further discoveries on the properties of dust in the coma of 67P/C-G are still expected, taking into account the enormous amount of data provided by the Rosetta mission. Finally, most results on cometary dust rely on flybys of 1P/Halley, indeed one HTC, and of numerous missions to JFCs. No mission to an OCC has taken place to date, while 1995 O1/Hale-Bopp, the only OCC extensively remotely observed, presented unexpected high linear polarization values.

We have presented updated results on the modeling of the dust continuum emission at sub-millimeter wavelengths of the protoplanetary disk around MWC 758. This modeling assumes that dust is comprised of moderately porous spherical particles, instead of (more conventional) compact spherical particles. The results presented in this paper focus on dust scattering, and show that using the same scattering phase function as that of the dust coma of comet 67P/C-G in the near-infrared, which includes both backward and forward scattering, leads to a better agreement between synthetic and observed maps of emission at sub-millimeter wavelengths. More work is needed to assess the robustness of this preliminary result, in particular to what extent a scattering phase function that is constrained by near-infrared observations can be applied at sub-millimeter wavelengths. In that regard, it would be interesting to examine the impact of the scattering phase function on our predictions at near-infrared wavelengths. Also, as already emphasized in Sect. 4.2, it seems very relevant to go beyond spherical particles in the modeling of the dust's dynamics and emission.

Our high-angular resolution observations of debris disks have now shown that light scattering properties of dust in debris disks are far from those predicted by Mie theory and in favor of irregular dust particles, as pointed out essentially by phase functions analysis. These results can greatly benefit from in situ and remote cometary observations and from advanced modeling of cometary dust, and at the same time provide examples of dust properties in stellar systems to put our Solar System into context. Two caveats however exist. On the one hand, the process releasing dust in debris disks (collisions between icy planetesimals) and in comets (dust sublimation) is different. Depending on the size of the planetesimals and the velocities of the impacts, released particles in debris disks may indeed present compact as well as fluffy structures. On the other hand, the number of resolved debris disks with sufficient signal-to-noise to extract the polarized and unpolarized phase function is still very small. The diversity of scattering behaviors is therefore an on-going area of investigation. From observations, the presence of silicates and organics has been shown (Hughes et al. 2018), which may render them relatively akin to the zodiacal cloud.

## 5. Conclusions and Perspectives

Tiny meteoroids entering the terrestrial atmosphere mostly originate in dust particles and debris released by nuclei of Jupiter-family comets. Independent approaches, relying on dynamical properties of interplanetary dust, on its thermal emission, or on comparison with a model combining the properties of cosmic dust with clues in the terrestrial atmosphere and at the South Pole, lead to this same conclusion.

The near-Earth properties of interplanetary dust, as derived from our studies of zodiacal light, fairly agree with those of cometary dust revealed by the Rosetta mission.

Extensive investigations of the dust particles in the coma of 67P/C-G have established that they consist of porous irregular and hierarchical aggregates, and that they are rich in organics, dominated by high molecular-weight components.

Such results ascertain the cometary origin of chondritic-porous IDPs collected in the Earth's stratosphere and of ultra carbonaceous Antarctica micrometeorites. Studies on tiny meteoroids can thus certainly benefit from advances in our understanding of the properties of cometary dust, and reciprocally.

Of major importance is the fact that the properties of cometary dust particles lead us, back in time, to a better understanding of the formation of comets, from accretion at low collision velocities of dust particles in the external regions of the protosolar nebula, and of the early evolution of the Solar System. Similarly, beyond in space, detailed studies of the properties of cometary and interplanetary dust in the Solar System lead us tentatively improve the modeling observations of light scattering by dust media in protoplanetary and debris disks. As an example, and without generalizing the properties of dust particles in circumstellar disks, it is already suggested that more realistic models going beyond compact spherical dust particles would be relevant.

Future space missions to comets, such as the ESA Comet Interceptor project (to tentatively flyby an OCC), together with light scattering observations from new telescopes, such as the Vera C. Rubin Observatory (previously referred to as LSST), should contribute to major progress in these domains. Meanwhile, studies aiming at extracting the dependence upon the phase angle for more debris disks and better characterizing a few individual systems are being carried out with current high-contrast imagers. Their goal should be to understand the role of the stellar properties or the orbital distance, and to allow more accurate comparisons with the properties of cometary and zodiacal dust, and thus links back to micrometeoroids, as studied in the Solar System.

**Acknowledgements**
This work was supported by the Programme National de Planétologie (PNP) of CNRS/INSU, co-funded by Centre National d'Études Spatiales (CNES). A.C.L.R. and J.L. acknowledge support from CNES in the realization of instruments devoted to space exploration of comets and in their scientific analysis, and Dr Hadamcik for her laboratory measurements on interplanetary dust analogs.